\documentclass[trackchanges]{aastex701}

\begin{document}

\title{Active galactic nuclei are not responsible for systematics in the empirical properties of type Ia supernovae}

\author[orcid=0000-0002-6576-7400]{Peter Clark}
\affiliation{School of Physics and Astronomy, University of Southampton, Southampton, SO17 1BJ, UK}
\email[show]{peterclark.astro@gmail.com}

\begin{abstract}

Despite being key cosmological anchors, the empirical properties of Type Ia supernovae (SNe Ia) remain subject to significant systematic uncertainties, the largest of which being their astrophysics and how this is linked to their environment. The relative importance of these uncertainties will grow as larger observational samples reduce statistical uncertainties. Here I explore if the presence of an active galactic nucleus (AGN) within a portion of SN Ia host galaxies could be one such systematic uncertainty. With the ZTF SNe Ia DR2 sample I find that, in the low redshift regime (z~$<$~0.15), the presence of AGNs in some SN Ia hosts does not produce a significant systematic effect on their measured Hubble residuals. 

\end{abstract}



\section{Introduction}
SNe Ia from different environments have systematically different distributions of properties that persist after standardization as cosmological distance indicators \citep[e.g.,][]{sullivan_2010_dependenceTypeIa,roman_2018_DependenceTypeIa, rigault_2020_StrongdependenceType, kelsey_2023_Concerningcoloureffect}. I investigate whether the presence of an AGN (a central supermassive black hole with a high accretion rate) in some SN Ia hosts influences this, either through direct effects on progenitor properties or by biasing measured host properties, such as stellar mass.

AGNs are known to affect their hosts via feedback (e.g., \citealt{fabian_2012_ObservationalEvidenceActive}), by channeling energy outwards and thus altering gas content and temperature. By comparing SNe Ia in galaxies with and without AGN, I explore whether these processes influence measured SN Ia properties and the cosmology inferred from them.

I use the distributions of \lq Hubble residuals\rq\ ($\Delta\mu$) between SN Ia samples as the diagnostic tool, i.e.,
\begin{equation}
    \Delta\mu = \mu_{obs} - \mu_{cosmo}.
\end{equation}
$\Delta\mu$ represents the difference between measured SN Ia brightness (following standardization) and the expected brightness from cosmological models. While differences in $\left\langle \Delta\mu \right\rangle$ of populations from different environments are well established, the specifics on how these relate to the physical conditions of those environments is debated. 

\section{Methods}
Data Release 2 (DR2) of the Zwicky Transient Facility's (ZTF) spectroscopically-classified SN Ia sample contains 3628 SNe Ia at redshifts $<$~0.3 \citep{rigault_2025_ZTFSNIa}. I explored two methods of identifying SNe in galaxies hosting an AGN; mid-infrared (MIR) color selection with data obtained from \textit{WISE} \citep{wright_2010_WIDEFIELDINFRAREDSURVEY, mainzer_2014_INITIALPERFORMANCENEOWISE}, and a cross-match with the MILLIQUAS AGN database \citep{flesch_2023_MillionQuasarsMilliquas}, which uses multiple diagnostics. 

Standard selection criteria matching \cite{rigault_2025_ZTFSNIa} are used, along with limiting to 0.01~$\leq~$z~$\leq$~0.15 to improve completeness and visual inspection to remove sources suffering from potential contamination or host-confusion, with the resulting sample consisting of 2425 SNe Ia.

MIR-AGN selection used the color cuts of \citet{stern_2012_MIDINFRAREDSELECTIONACTIVE}, \citet{mateos_2012_UsingBrightUltrahard} and \cite{assef_2013_MidinfraredSelectionActive} with data retrieved from the NASA/IPAC infrared science archive \citep{wiseteam_2020_NEOWISE2BandPostCryo} and processed following \cite{clark_2024_Longtermfollowupobservations}. The probability of each galaxy hosting an AGN was determined per epoch via Monte Carlo analysis, then averaged to provide an overall probability. Following cuts on apparent magnitude (\textit{W}1~$<$~14.70, \textit{W}2~$<$~14.45), and further contamination/confusion checks, insufficient AGN (between 2 and 10) were identified by any diagnostic at a probability exceeding 70\% to be useful for comparison. This is consistent with previous analyses such as \citet{sartori_2015_searchactiveblack}, who classified 0.4\% of their local galaxies as MIR-AGN hosts.

Following crossmatching with MILLIQUAS (within 3$''$), 58 AGN-hosting galaxies (3.5\%) are identified, with this sample adopted for further analysis.

\section{Results}

\citet{rigault_2025_ZTFSNIa} indicates the published ZTF light curves and associated parameters are not suitable for precise cosmological inference. However, here I am interested if the presence of AGNs in some SNe hosts is a source of systematic uncertainty, rather than directly quantifying the underlying cosmology. I therefore adopt the model of \citet{ginolin_2025_ZTFSNIa} i.e., $\alpha$ = 0.161~$\pm$~0.010 and $\beta$ = 3.05~$\pm$~0.060 and set M$_{0}$ to -19.4 (see Eqn~\ref{Eqn_Tripp}), noting our samples have somewhat differing selection criteria.

\begin{equation}
\label{Eqn_Tripp}
    \mu_{obs} = -2.5log_{10}(x_{0}) + M_{0} + (\alpha x_{1}) - (\beta c)
\end{equation}

\begin{figure*}
    \centering
    \includegraphics[width=0.87\linewidth]{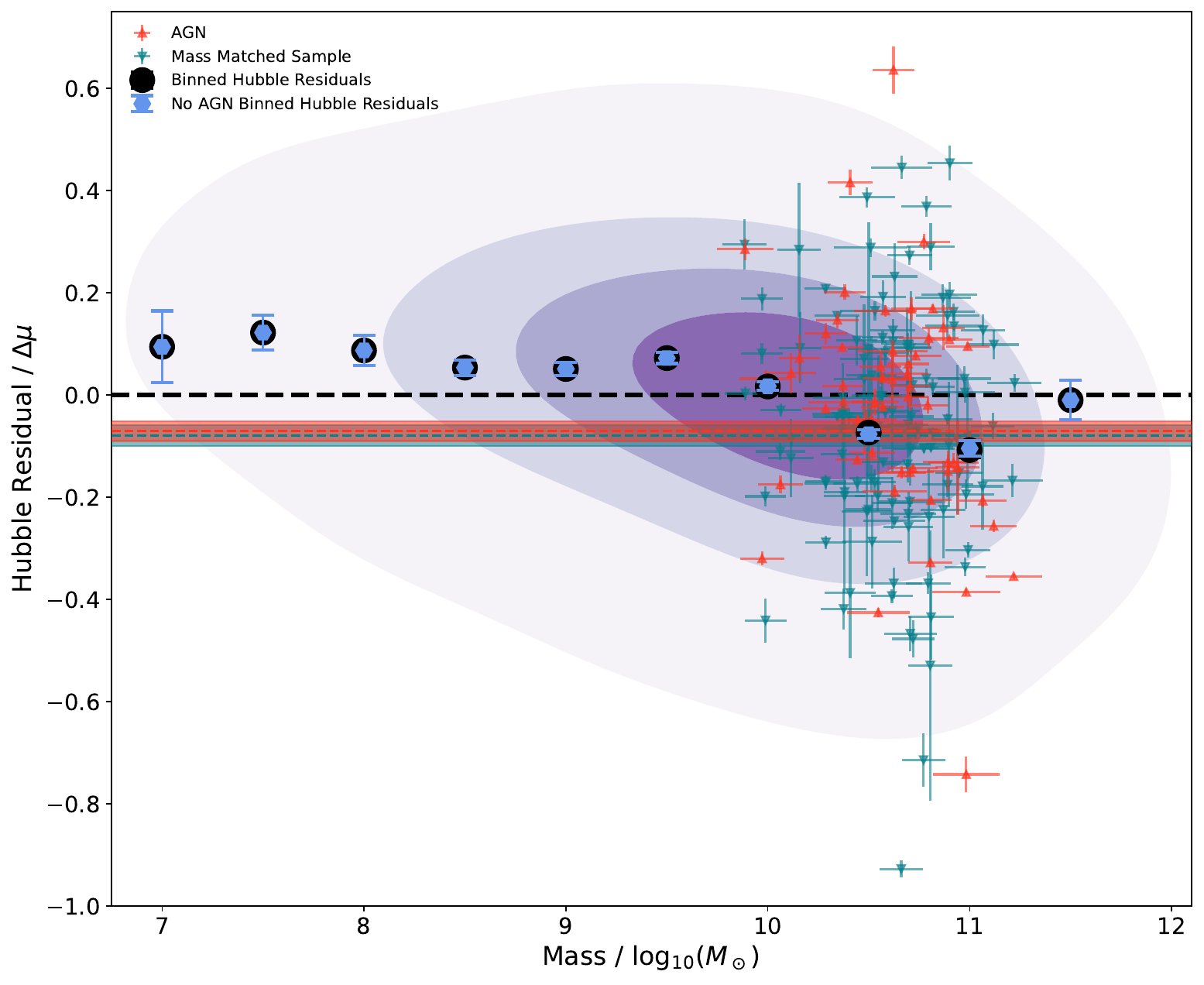}
    \caption{The $\Delta\mu$ distribution of the full SNe Ia sample versus host mass (purple shading). Red upward triangles mark SNe in AGN-hosting galaxies, while teal downward triangles show mass-matched non-AGN counterparts. Weighted means (dashed lines) agree within uncertainties. Full-sample means are shown as circles and hexagons.}
    \label{fig1}
\end{figure*}

I determine $\mu_{cosmo}$ using the flat $\Lambda$CDM model of \citealt{plankcollaboration_2020_Planck2018results}, and normalize so $\left\langle \Delta\mu \right\rangle$ of the full sample is zero. I then compare the $\Delta\mu$ distribution of those SNe in galaxies hosting AGN to both the overall sample's distribution, and a mass-matched sample of the unique galaxies closest in mass both above and below each AGN hosting galaxy, see Fig.~\ref{fig1}.

All but three galaxies hosting AGN have masses exceeding $10^{10}M_{\odot}$ and would be considered high-mass galaxies in SN Ia cosmology studies. These are known to host systemically brighter SNe Ia (those with negative $\Delta\mu$) which is retrieved here. However, when the AGN-containing and non-AGN mass-matched samples are compared, there is no significant difference between the $\left\langle \Delta\mu \right\rangle$: -0.07~$\pm$~0.02 for the AGN sample, versus -0.08~$\pm$~0.02 for the non-AGN mass-matched sample.

Therefore, the presence of AGNs in a subpopulation of SN Ia hosts cannot account for the observed difference in the brightnesses measured for SNe Ia in high-mass galaxies, at least in this low redshift regime.

\section{Conclusions}

The ZTF DR2 SNe Ia sample offers a valuable opportunity to study low-redshift SNe Ia subpopulations. However, too few reliably classifiable MIR-AGN hosts within it to perform robust analysis. However, the small number present indicates that this subpopulation cannot drive the observed differences across SNe Ia. SNe Ia hosts with AGNs identified through broader diagnostics form a small, but testable population (3.5~\%). A comparison of this sub-population's $\left\langle \Delta\mu \right\rangle$ to a mass-matched sample of non-AGN containing SNe Ia hosts finds no statistically significant difference.

I conclude that the presence of AGNs in some SN Ia hosting galaxies is unlikely to be responsible for significant systematic trends or differences in observed SN Ia properties in the low-redshift regime (z~$<$~0.15).

Expansion of this work with future samples \citep[e.g., 4MOST TiDES;][]{frohmaier_2025_TiDES4MOSTTime} and employing data from the Spectro-Photometer for the History of the Universe, Epoch of Reionization and Ices Explorer \citep[SPHEREx;][]{bock_2026_SPHERExSatelliteMission} is planned.

\begin{acknowledgments}
PC was supported by the Science \& Technology Facilities Council grant ST/Y001850/1.

I thank the organizers of the ‘Cosmic Lighthouses’ conference for inspiring this work, and Lisa Kelsey, Mark Sullivan, and Brodie Popovic for insightful discussions.
\end{acknowledgments}




%
\facilities{WISE, ZTF}

\software{astropy \citep{astropycollaboration_2022_AstropyProjectSustaining}}

\bibliography{MyLibrary}{}

@article{assef_2013_MidinfraredSelectionActive,
  title = {Mid-Infrared {{Selection}} of {{Active Galactic Nuclei}} with the {{Wide-field Infrared Survey Explorer}}. {{II}}. {{Properties}} of {{WISE-selected Active Galactic Nuclei}} in the {{NDWFS Bo\"otes Field}}},
  author = {Assef, R. J. and Stern, D. and Kochanek, C. S. and Blain, A. W. and Brodwin, M. and Brown, M. J. I. and Donoso, E. and Eisenhardt, P. R. M. and Jannuzi, B. T. and Jarrett, T. H. and Stanford, S. A. and Tsai, C. -W. and Wu, J. and Yan, L.},
  year = 2013,
  month = jul,
  journal = {ApJ},
  volume = {772},
  pages = {26},
  issn = {0004-637X},
  doi = {10.1088/0004-637X/772/1/26},
  urldate = {2023-03-13}
}

@article{astropycollaboration_2022_AstropyProjectSustaining,
  title = {The {{Astropy Project}}: {{Sustaining}} and {{Growing}} a {{Community-oriented Open-source Project}} and the {{Latest Major Release}} (v5.0) of the {{Core Package}}},
  shorttitle = {The {{Astropy Project}}},
  author = {{Astropy Collaboration} and {Price-Whelan}, Adrian M. and Lim, Pey Lian and Earl, Nicholas and Starkman, Nathaniel and Bradley, Larry and Shupe, David L. and Patil, Aarya A. and Corrales, Lia and Brasseur, C. E. and N{\"o}the, Maximilian and Donath, Axel and Tollerud, Erik and Morris, Brett M. and Ginsburg, Adam and Vaher, Eero and Weaver, Benjamin A. and Tocknell, James and Jamieson, William and {van Kerkwijk}, Marten H. and Robitaille, Thomas P. and Merry, Bruce and Bachetti, Matteo and G{\"u}nther, H. Moritz and Aldcroft, Thomas L. and {Alvarado-Montes}, Jaime A. and Archibald, Anne M. and B{\'o}di, Attila and Bapat, Shreyas and Barentsen, Geert and Baz{\'a}n, Juanjo and Biswas, Manish and Boquien, M{\'e}d{\'e}ric and Burke, D. J. and Cara, Daria and Cara, Mihai and Conroy, Kyle E. and Conseil, Simon and Craig, Matthew W. and Cross, Robert M. and Cruz, Kelle L. and D'Eugenio, Francesco and Dencheva, Nadia and Devillepoix, Hadrien A. R. and Dietrich, J{\"o}rg P. and Eigenbrot, Arthur Davis and Erben, Thomas and Ferreira, Leonardo and {Foreman-Mackey}, Daniel and Fox, Ryan and Freij, Nabil and Garg, Suyog and Geda, Robel and Glattly, Lauren and Gondhalekar, Yash and Gordon, Karl D. and Grant, David and Greenfield, Perry and Groener, Austen M. and Guest, Steve and Gurovich, Sebastian and Handberg, Rasmus and Hart, Akeem and {Hatfield-Dodds}, Zac and Homeier, Derek and Hosseinzadeh, Griffin and Jenness, Tim and Jones, Craig K. and Joseph, Prajwel and Kalmbach, J. Bryce and Karamehmetoglu, Emir and Ka{\l}uszy{\'n}ski, Miko{\l}aj and Kelley, Michael S. P. and Kern, Nicholas and Kerzendorf, Wolfgang E. and Koch, Eric W. and Kulumani, Shankar and Lee, Antony and Ly, Chun and Ma, Zhiyuan and MacBride, Conor and Maljaars, Jakob M. and Muna, Demitri and Murphy, N. A. and Norman, Henrik and O'Steen, Richard and Oman, Kyle A. and Pacifici, Camilla and Pascual, Sergio and {Pascual-Granado}, J. and Patil, Rohit R. and Perren, Gabriel I. and Pickering, Timothy E. and Rastogi, Tanuj and Roulston, Benjamin R. and Ryan, Daniel F. and Rykoff, Eli S. and Sabater, Jose and Sakurikar, Parikshit and Salgado, Jes{\'u}s and Sanghi, Aniket and Saunders, Nicholas and Savchenko, Volodymyr and Schwardt, Ludwig and {Seifert-Eckert}, Michael and Shih, Albert Y. and Jain, Anany Shrey and Shukla, Gyanendra and Sick, Jonathan and Simpson, Chris and Singanamalla, Sudheesh and Singer, Leo P. and Singhal, Jaladh and Sinha, Manodeep and Sip{\H o}cz, Brigitta M. and Spitler, Lee R. and Stansby, David and Streicher, Ole and {\v S}umak, Jani and Swinbank, John D. and Taranu, Dan S. and Tewary, Nikita and Tremblay, Grant R. and {de Val-Borro}, Miguel and Van Kooten, Samuel J. and Vasovi{\'c}, Zlatan and Verma, Shresth and {de Miranda Cardoso}, Jos{\'e} Vin{\'i}cius and Williams, Peter K. G. and Wilson, Tom J. and Winkel, Benjamin and {Wood-Vasey}, W. M. and Xue, Rui and Yoachim, Peter and Zhang, Chen and Zonca, Andrea and {Astropy Project Contributors}},
  year = 2022,
  month = aug,
  journal = {ApJ},
  volume = {935},
  pages = {167},
  publisher = {IOP},
  issn = {0004-637X},
  doi = {10.3847/1538-4357/ac7c74},
  urldate = {2024-06-18}
}

@article{bock_2026_SPHERExSatelliteMission,
  title = {The {{SPHEREx Satellite Mission}}},
  author = {Bock, James J. and Aboobaker, Asad M. and Adamo, Joseph and Akeson, Rachel and Alred, John M. and Alibay, Farah and Ashby, Matthew L. N. and Bach, Yoonsoo P. and Bleem, Lindsey E. and Bolton, Douglas and Braun, David F. and Bruton, Sean and Bryan, Sean A. and Chang, Tzu-Ching and Chen, Shuang-Shuang and Cheng, Yun-Ting and Cheshire, IV, James R. and Chiang, Yi-Kuan and {Choppin de Janvry}, Jean and Condon, Samuel and Cook, Walter R. and Cooray, Asantha and Crill, Brendan P. and Cukierman, Ari J. and Dor{\'e}, Olivier and Dowell, C. Darren and {Dubois-Felsmann}, Gregory P. and Eifler, Tim and Everett, Spencer and Fabinsky, Beth E. and Faisst, Andreas L. and Fanson, James L. and Farrington, Allen H. and Fatahi, Tamim and Fazar, Candice M. and Feder, Richard M. and Frater, Eric H. and Grasshorn Gebhardt, Henry S. and Giri, Utkarsh and Goldina, Tatiana and Gorjian, Varoujan and Habib, Salman and Hart, William G. and Heinrich, Chen and Hora, Joseph L. and Huai, Zhaoyu and Hui, Howard and Jo, Young-Soo and Jeong, Woong-Seob and Kang, Jae Hwan and Kang, Miju and Kecman, Branislav and Kim, Chul-Hwan and Kim, Jaeyeong and Kim, Minjin and Kim, Young-Jun and Kim, Yongjung and Kirkpatrick, J. Davy and Kobayashi, Yosuke and Korngut, Phil M. and Krause, Elisabeth and Lee, Bomee and Lee, Ho-Gyu and Lee, Jae-Joon and Lee, Jeong-Eun and Lisse, Carey M. and Mariani, Giacomo and Masters, Daniel C. and Mauskopf, Philip D. and Melnick, Gary J. and Minasyan, Mary H. and Mirocha, Jordan and Miyasaka, Hiromasa and Moore, Anne and Moore, Bradley D. and Murgia, Giulia and Naylor, Bret J. and Nelson, Christina and Nguyen, Chi H. and Nguyen, Hien T. and Noh, Jinyoung K. and Padin, Stephen and Paladini, Roberta and Park, Sung-Joon and Penanen, Konstantin I. and Putnam, Dustin S. and Pyo, Jeonghyun and Ramachandra, Nesar and Ramanathan, Keshav and Rustamkulov, Zafar and Reiley, Daniel J. and Rice, Eric B. and Rocca, Jennifer M. and Seok, Ji Yeon and Smith, Roger and Stober, Jeremy and Susca, Sara and Teplitz, Harry I. and Thelen, Michael P. and Tolls, Volker and Torrini, Gabriela and Trangsrud, Amy R. and Unwin, Stephen and Velicheti, Phani and Wang, Pao-Yu and Wen, Robin Y. and Werner, Michael W. and Williams, Abby E. and Williamson, Ross and Wincentsen, James and Windhorst, Rogier A. and Yang, Soung-Chul and Yang, Yujin and Zemcov, Michael},
  year = 2026,
  month = mar,
  journal = {ApJ},
  volume = {999},
  pages = {139},
  publisher = {IOP},
  issn = {0004-637X},
  doi = {10.3847/1538-4357/ae2be2},
  urldate = {2026-06-30}
}

@article{clark_2024_Longtermfollowupobservations,
  title = {Long-Term Follow-up Observations of Extreme Coronal Line Emitting Galaxies},
  author = {Clark, Peter and Graur, Or and Callow, Joseph and Aguilar, Jessica and Ahlen, Steven and Anderson, Joseph P. and Berger, Edo and {M{\"u}ller-Bravo}, Tom{\'a}s E. and Brink, Thomas G. and Brooks, David and Chen, Ting-Wan and Claybaugh, Todd and {de la Macorra}, Axel and Doel, Peter and Filippenko, Alexei V. and {Forero-Romero}, Jamie E. and Gomez, Sebastian and Gromadzki, Mariusz and Honscheid, Klaus and Inserra, Cosimo and Kisner, Theodore and Landriau, Martin and Makrygianni, Lydia and Manera, Marc and Meisner, Aaron and Miquel, Ramon and Moustakas, John and Nicholl, Matt and Nie, Jundan and Onori, Francesca and Palmese, Antonella and Poppett, Claire and Reynolds, Thomas and Rezaie, Mehdi and Rossi, Graziano and Sanchez, Eusebio and Schubnell, Michael and Tarl{\'e}, Gregory and Weaver, Benjamin A. and Wevers, Thomas and Young, David R. and Zheng, WeiKang and Zhou, Zhimin},
  year = 2024,
  month = feb,
  journal = {MNRAS},
  volume = {528},
  number = {4},
  pages = {7076--7102},
  issn = {0035-8711},
  doi = {10.1093/mnras/stae460},
  urldate = {2024-02-26}
}

@article{fabian_2012_ObservationalEvidenceActive,
  title = {Observational {{Evidence}} of {{Active Galactic Nuclei Feedback}}},
  author = {Fabian, A. C.},
  year = 2012,
  month = sep,
  journal = {ARA\&A},
  volume = {50},
  pages = {455--489},
  issn = {0066-4146},
  doi = {10.1146/annurev-astro-081811-125521},
  urldate = {2026-01-29}
}

@misc{wiseteam_2020_NEOWISE2BandPostCryo,
  title = {{{NEOWISE}} 2-{{Band Post-Cryo Single Exposure}} ({{L1b}}) {{Source Table}}},
  author = {{WISE Team}},
  year = 2020,
  publisher = {IPAC},
  doi = {10.26131/IRSA124},
  urldate = {2026-07-01}
}
\bibliographystyle{aasjournalv7}



\end{document}